# Large and non-linear permeability amplification with polymeric additives in hydrogel membranes

Malak Alaa Eddine[1,2], Sabrina Belbekhouche*[2], Sixtine de Chateauneuf-Randon[1], Thomas Salez[3], Artem Kovalenko[1], Bruno Bresson[1], Cécile Monteux*[1]

1- Laboratoire Sciences et Ingénierie de la Matière Molle, ESPCI Paris, 10 rue Vauquelin, Cedex 05 75231 Paris, France
2- Université Paris Est Créteil, CNRS, Institut Chimie et Matériaux Paris Est, UMR 7182, 2 Rue Henri Dunant, 94320 Thiais, France
3- Univ. Bordeaux, CNRS, LOMA, UMR 5798, F-33400 Talence, France.

* Authors for correspondence:
E-mail addresses: Cecile.monteux@espci.fr (C. Monteux).
belbekhouche@icmpe.cnrs.fr (S. Belbekhouche).

**Abstract**

Hydrogels which are hydrophilic and porous materials have recently emerged as promising systems for filtration applications. In our study, we prepare hydrogel membranes by the photopolymerization of a mixture of poly (ethylene glycol) diacrylate (PEGDA) and large poly(ethylene glycol) (PEG) chains of 300 000 g.mol$^{-1}$ in the presence of a photoinitiator. We find that this addition of free PEG chains induces a large and non-linear increase of the water permeability. Indeed, by changing the content of PEG chains added, we obtain variations of the hydrogel water permeability over two orders of magnitude. The highest water permeability values are obtained for the membranes when the PEG concentration is equal to its critical overlap concentration $C^*$. Moreover, we find that the flow rate of water through the membranes varies non-linearly with the pressure. We relate this result to the deformability of the membranes as the applied pressure leads to a compression of the pores. This study provides new perspectives for the design of flexible hydrogel membranes with controlled permeability and their application in water treatment and bioseparation.

# 1. Introduction

Pressure-driven membrane technologies, such as microfiltration, ultrafiltration, nanofiltration and reverse osmosis, have proven their effectiveness in a broad range of water treatment applications [1, 2]. The porous structure of membranes, either composed of ceramics or polymers, controls the size of the particles that permeates through the membrane. Fouling of membranes with bio-organisms and proteins usually limits the life time and permeability of membranes [3, 4].

Hydrogels, which are networks of polymer chains in water, have a porous and hydrophilic structure which has recently attracted the attention of researchers in the context of filtration and separation [5]. Their network structure also enables ones to control the Brownian diffusion of species through them [6, 7] which can be used for drug delivery purpose [8]. Controlling the transport of solvents, ions, solutes and particles in such polymer networks is also desirable for other applications such as catalysis, fuel cells and batteries and can be done by controlling their microscopic morphology and porosity [9]. In the context of water treatment [10], biomolecular separation [11, 12], virus filtration [13] or even crude oil emulsion separation [14], thin coatings of hydrogels deposited on conventional filtration membranes have been used to prevent fouling of membranes by proteins. Indeed, classical hydrophobic membranes made of polysulfone (PSF) [15], polyethersulfone (PES) [16], or polyvinylidenefluoride (PVDF) [17], and coated with a hydrogel layer enable both to increase the hydrophilicity of the membranes and to decrease their affinity with hydrophobic proteins.

Recently poly (ethylene glycol) diacrylate (PEGDA)-based hydrogels have been used either as thin coatings on hydrophobic membranes [15, 18] or as stand-alone thick membranes [19, 20]. Ju and colleagues [19] studied the water permeability of a family of cross-linked poly (ethylene glycol) diacrylate (PEGDA) membranes. PEGDA is particularly suitable for filtration experiments as it can sustain high filtration pressures owed to its excellent mechanical properties [21].

Micron-sized PEGDA/PEG hydrogels are also used for microfluidic filtration applications [22-24] where small free PEG chains ($\overline{Mw}$ ~1000 g.mol$^{-1}$) are added in the prepolymerization mixture in the context of protein crystallization [25]. After polymerization, the small free PEG chains are expected to be rinsed out of the membranes hence enabling the increase of the permeability of PEGDA hydrogels over one order of magnitude depending on the PEG concentration [24]. To control the permeability of PEGDA hydrogels, other studies report the polymerization of PEGDA in the presence of a sacrificial template (or porogens) such as salt crystals [26], or sacrificial particles [27] which are dissolved after polymerization. Although these methods allow the formation of larger porosities in the hydrogels, some of them involve removing these templates using a chemical treatment which limits their application.

In our study, we focus on the filtration properties of free-standing hydrogel membranes, of millimetric thickness, composed of a PEGDA matrix in which large free PEG chains ($\overline{Mw}$=300 000 g.mol$^{-1}$) are added at varying concentrations. Here, the free PEG chains are not used as templates but remain trapped inside the PEGDA matrix. Hence, a great advantage of these membranes is that they can be prepared in a one-step process. We show that varying the PEG concentration enables tuning the PEG/PEGDA hydrogel's permeability over two orders of magnitude, including an optimum with the PEG concentration corresponding to the critical overlap concentration of PEG solutions. Moreover, we show that the permeability of the PEG/PEGDA membranes strongly depends on the applied pressure. We suggest that this non-linear behavior is due to a deformation of the hydrogel matrix induced by the confined viscous flow. These results are promising for the development of versatile macroporous hydrogel membranes with tunable and non-linear water permeabilities.

## 2. Experimental section

### 2.1. Materials

We use poly (ethylene glycol) diacrylate PEGDA ($\overline{Mw}$=700 g.mol$^{-1}$) oligomers with 13 ethylene oxide units and 4-(2-hydroxyethoxy) phenyl 2-hydroxy-2-propyl ketone (Irgacure 2959) photoinitiator which are purchased from Sigma–Aldrich. Linear poly (ethylene glycol) (PEG) ($\overline{Mw}$ =300 000 g.mol$^{-1}$, Đ =2.1) is purchased from Serva. Water is purified with a Milli-Q reagent system (Millipore).

### 2.2. PEGDA hydrogels preparation

The PEGDA and PEG/PEGDA membranes are synthesized via UV-initiated free-radical photopolymerization, using Irgacure 2959 as the photoinitiator. The prepolymerization solution is prepared by adding 0.1 wt% photoinitiator (Irgacure 2959) into pure PEGDA. After stirring, the solution is mixed with water to obtain a prepolymerization solution composed of 16 wt% PEGDA and 84 wt% water. The prepolymerization solution is then sandwiched between two glass plates (120x80 mm$^2$) which are separated by 1-mm-thick spacers to obtain a membrane thickness of 1 mm. Then the solution is polymerized under irradiation with UV light (Intensity =1800 μW/cm$^2$) with a wavelength of 365 nm for 10 min. After polymerization, the obtained hydrogels are placed in a Petri dish with pure water for at least 24 hours prior to filtration experiments, in order to remove any unreacted PEGDA oligomers or unentrapped PEG chains. As explained in SI 1, the obtained hydrogels right after polymerization contain 82% of water and 18% of PEGDA. Moreover, once immersed in deionized water, they do not swell or deswell over four days at room temperature, consistently with the study of Ju *et al.*[19].

To obtain the PEG/PEGDA hydrogels, PEG-300 000 g.mol$^{-1}$ is dissolved in the prepolymerization solution. To prepare the sample, we keep the masses of PEGDA and water constant and add varying quantities of PEG-300 000 g.mol$^{-1}$ in the prepolymerization solution so that the PEG weight percentage $\Phi_{PEG}$ in the prepolymerization solution ranges between 0.4 wt% and 7.7 wt%. For the filtration and mechanical measurements, the membranes are then cut to obtain 1-mm-thick disks of diameter either 14 mm or 45 mm using punch cutters with corresponding diameters. As shown in SI 1, by weighing the hydrogel disks in the preparation state, and after drying, we find that the water content is comprised between 80 and 82% for all the prepared hydrogels. Moreover, no significant swelling is measured when the hydrogels are immersed in pure water at room temperature.

### 2.3. Chemical composition of hydrogels

The chemical composition was analyzed by Fourier-transform infrared spectroscopy FTIR (Bruker, Tensor 27 instrument equipped with a Digitect DLaTGS detector). The FTIR spectroscopy resolution is about 4 cm$^{-1}$ and the infrared radiation ranging approximately from 4000 to 400 cm$^{-1}$. The number of averaged scans is 32.

### 2.4. Atomic Force Microscopy (AFM) measurements

AFM images were obtained with a Bruker Icon microscope driven by a Nanoscope V controller. The surface of the hydrogel membrane immersed in water was observed in Peak Force mode. The height images were acquired with a cantilever of spring constant 0.7 N.m$^{-1}$ specially designed for this application. In this mode, similar to a rapid approach-retract experiment, the cantilever oscillates at a frequency of 1 kHz. The scanning frequency was 0.7 Hz and the maximum force was set to 500 pN.

### 2.5. Mechanical measurements

Disc-shaped hydrogel membranes were used to test the mechanical properties. Before mechanical testing, the gels were conserved in water overnight. Discs were 14 mm in diameter and ~1 mm in thickness. Samples were tested under compression using an Instron 5565 testing machine at a deformation speed of 0.01 mm.s$^{-1}$. Discs were compressed between 10 and 40% of their original thickness and then unloaded to determine their elastic recoil capacity. The stress σ was calculated from the ratio between the force F and the initial sample area $S_0$ using the relation $\sigma=F/S_0$. The strain ε was calculated from the sample thickness in the initial state ($h_0$) and in the compressed state ($h$) as $\varepsilon = h/h_0 \times 100$ (%). The effective Young modulus of each gel was calculated from the slope of the stress–strain curve using the Hooke's law $E = \sigma/\varepsilon$ in the range of 2-4 % in strain, in the loading phase. For all samples, E was calculated from adjusting the stress-strain data at strain 0-4%, while some curve variability was observed at the very low strains (0-1%) because of parallelism issues.

### 2.6. Filtration experiments

Water permeability through conventional and PEG-modified PEGDA hydrogels was measured using a dead-end ultrafiltration UF cell obtained from Fisher Scientific S.A.S. (Amicon Model 8050, 50 mL for 45 mm Filters), as shown in Figure 1. The filtration was performed at ambient temperature, with Milli-Q water as the feed solution. A maximum feed pressure of 1 bar was used. The membrane with the area 15.90 cm$^2$ was fixed in the membrane holder of the cell. We first checked that the flow rate at a given pressure $\Delta P$ was constant during hours for all the samples. Then, standard filtration experiments were performed by increasing the pressure from 100 mbar to 1 bar and waiting 10 minutes at each pressure, while weighing the liquid permeate as a function of time with a balance (Sartorius) to obtain a precise measurement of the flow rate $Q$.

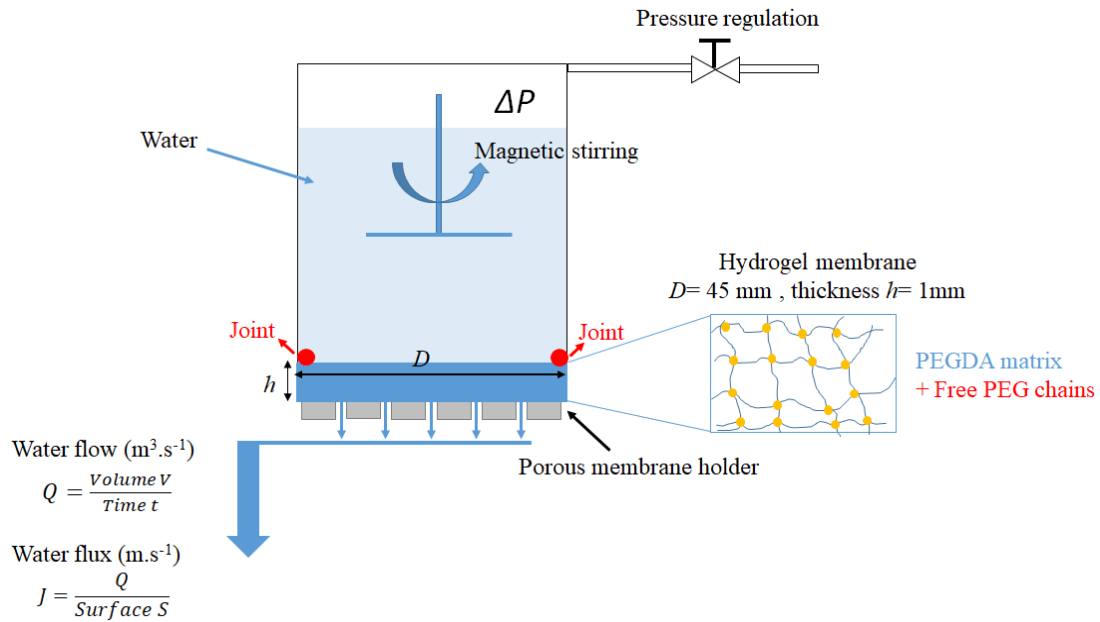

*Figure 1 : Representative schematic of the filtration experiments using an ultrafiltration stirred cell.*

Permeate flow rate $Q$ was recorded, and the water intrinsic permeability $K$ was calculated by the following equation (1):

$$K = \frac{Q\mu h}{\Delta P S} \quad (1)$$

where $Q$ is the water flow rate (m³.s⁻¹) calculated from the slope of the variation of the accumulated permeate volume (m³) as a function of time (s), $\mu$ is the water viscosity (Pa.s), $h$ is the hydrogel thickness (m), $S$ is the surface area of the hydrogel membrane (m²) and $\Delta P$ is the pressure difference across the membrane (Pa). The water flux is calculated from the value of water flow rate according to equation (2):

$$J = \frac{Q}{S} \quad (2)$$

Permeate samples were analyzed using a total organic carbon (TOC-L series from Shimadzu) in order to determine if PEG-300 000 g.mol⁻¹ chains were washed out of the gel during the filtration experiments.

## 3. Results

### 3.1. Hydrogel membranes characterization

In order to follow the polymerization reaction of PEGDA oligomers, FTIR spectra were acquired for the as-received PEGDA oligomer, conventional PEGDA sample and PEG-modified PEGDA samples. The analyzed samples were obtained from hydrogels via drying in a vacuum oven at 80° C for 2-3 h. Figure 2 presents the FTIR spectra of dried PEGDA hydrogels membranes prepared with 16 wt % of PEGDA and various PEG-300 000 g.mol⁻¹ contents in prepolymerization mixture.

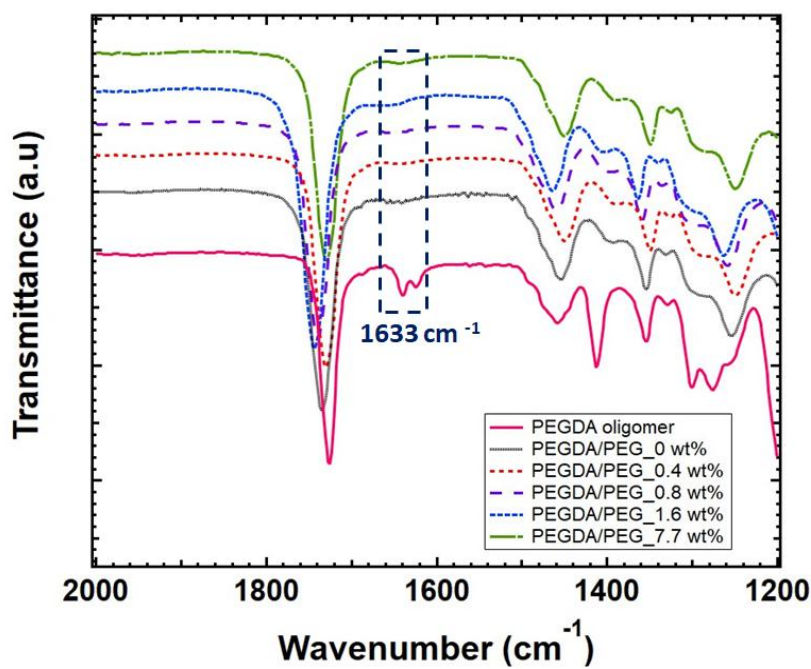

*Figure 2 : FTIR spectra of dried PEGDA hydrogels membranes prepared with 16 wt% of PEGDA and various PEG-300 000 g.mol$^{-1}$ contents in prepolymerization mixture compared to the spectrum of the as-received PEGDA oligomer.*

In the PEGDA oligomer spectrum, we find the characteristic peaks that are attributed to the carbonyl groups C=O at 1724 cm$^{-1}$ and the C=C bonds at 1633 cm$^{-1}$. For the cross-linked PEGDA or mixture of PEGDA/PEG samples, we observe that the peaks of carbonyl groups remain the same, while the characteristic peak of the C=C bonds at 1633 cm$^{-1}$ from PEGDA has disappeared. This suggests that the hydrogel films were successfully synthesized after UV crosslinking. Our results confirm that the PEGDA polymerization reaction is complete even after the addition of free PEG-300 000 g.mol$^{-1}$ chains of different concentrations.

While the membrane of PEGDA/PEG_0 wt% is colorless and transparent, the addition of free PEG chains induces a strong turbidity of the samples as shown in Figure 3 a, b, c consistently with a heterogeneous structure with spatial variations of the index of refraction that scatter light and which size ranges from several hundreds of nanometers to micron size.

To further characterize the hydrogels structure, AFM measurements were performed in water for a series of PEGDA hydrogel membranes prepared with various PEG-300 000 g.mol$^{-1}$ contents (Figure 3 d to i). For PEGDA/PEG_0 wt %, the hydrogel presents a surface with heterogeneities of the order of 100 to 200 nm in diameter which can be seen in Figure 3 d with a field of view of 20 μm and more precisely in Figure 3 g with a field of view of 5 μm. This is consistent with the work of Molina *et al.* who evidenced a structure with PEGDA rich zones coexisting with 200 nm voids filled with water [28]. The authors showed that the volume fraction of the water voids decreases when the PEGDA concentration increases.

For the PEGDA/PEG samples 0.8 wt% and 4 wt%, the AFM images with a field of view of 20 µm Figure 3 (e-f) show micron size heterogeneities consistently with the increase in turbidity observed upon addition of PEG in the PEGDA samples.

When imaging the 0.8 wt% PEG/PEGDA sample with a 5 µm field of view (Figure 3 h) we find that the micron sized zones contain voids of diameter of the order of 40 nm, which is smaller than the voids observed for pure PEGDA. Several 200 nm large voids can also be seen and seem to be located in the periphery of the micron size areas. From these results we suggest that a phase separation between PEGDA and PEG controls the heterogeneous structure of the samples. According to Molina *et al.*[28] who showed that the volume fraction of the voids reflects the PEGDA content we suggest that the micron sized zones that contain 40 nm large voids are enriched in PEGDA with respect to the areas at their periphery.

As the PEG content increases to 4 wt%, the micron sized heterogeneities can still be seen on the 20 µm field of view image (Figure 3 f). However, the nanometric voids can no longer be seen on the 5 µm field of view image (Figure 3 i) neither in the micron sized zones nor at their periphery. The absence of voids in the micron sized zones may be due to an enrichment in PEGDA of the PEGDA rich zones. We note that the absence of nanometric water voids may also be due to a strong increase of the PEG concentration, filling the water voids either in the PEGDA rich areas or at their periphery.

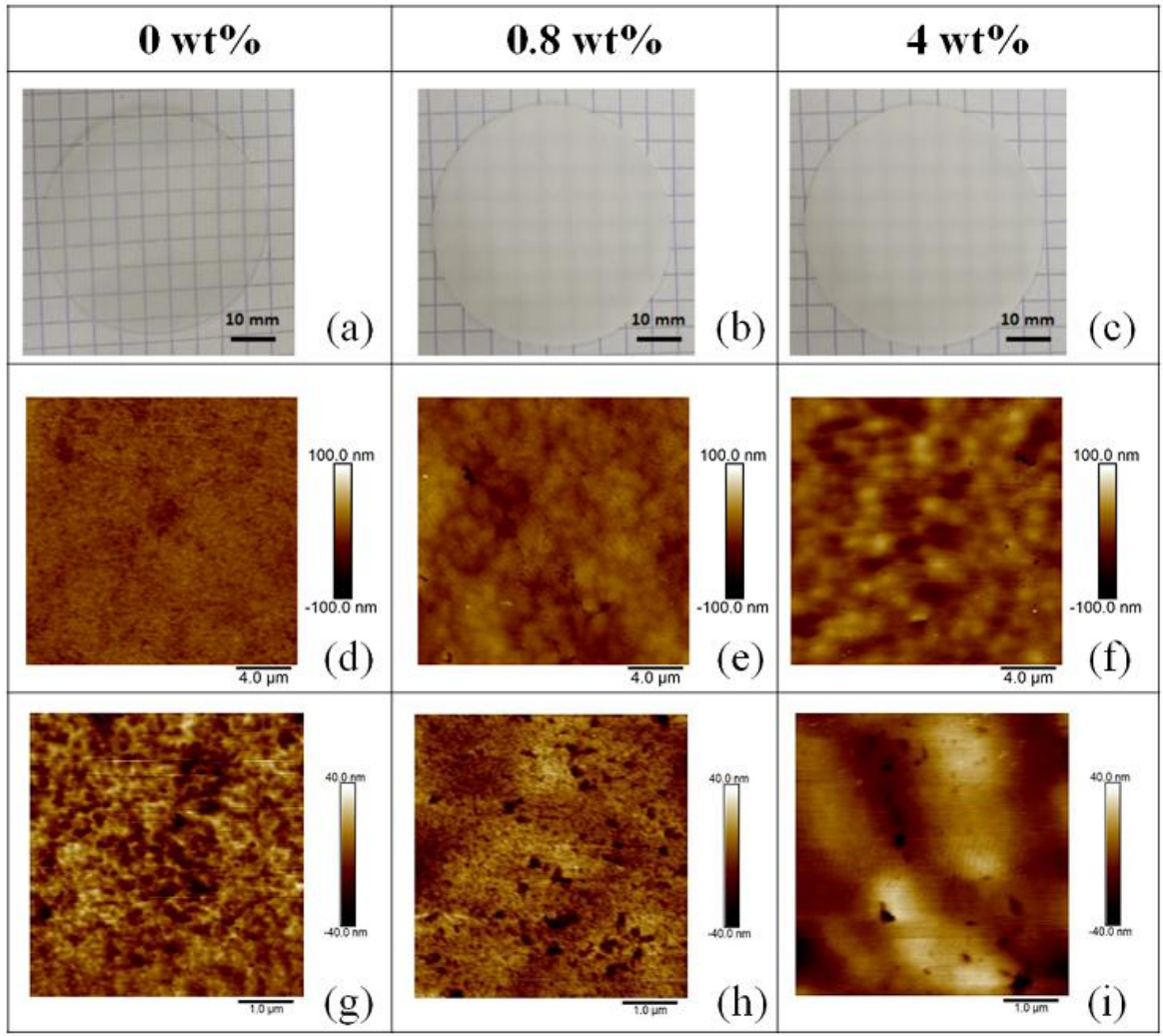

*Figure 3 : Photographs of PEGDA hydrogel membranes prepared with PEGDA and a) 0 wt%; b) 0.8 wt% and c) 4 wt% of PEG-300 000 g.mol$^{-1}$ in the prepolymerization mixture.*
*Surface AFM images with 20 µm of field of view of PEGDA hydrogel membranes prepared with PEGDA and d) 0 wt%; e) 0.8 wt% and f) 4 wt% of PEG-300 000 g.mol$^{-1}$ in the prepolymerization mixture.*
*Surface AFM images with 5 µm of field of view of PEGDA hydrogel membranes prepared with PEGDA and g) 0 wt%; h) 0.8 wt% and i) 4 wt% of PEG-300 000 g.mol$^{-1}$ in the prepolymerization mixture.*

### 3.2. Mechanical characterization

To further characterize the membranes samples, compression experiments were carried out for PEGDA/PEG hydrogel membranes prepared with various contents of PEG-300 000 g.mol$^{-1}$.

Figure 4 a presents the variation of the stress as a function of the strain in the linear regime (0-4 %) at a deformation rate of 0.01 mm.s$^{-1}$, for deformations between 0 and 4% and stresses below 0.05 MPa, which are relevant to the filtration experiments where pressures below 1 Bar (i.e. 0.1 MPa) are applied.

For PEGDA membranes prepared without PEG (0 wt%), the effective Young modulus value, calculated from the slope of the stress-strain curve at low strain (0-4%), is 1 MPa, which is of the same order of magnitude as the values reported in the literature [29, 30]. The addition of free PEG chains, from 0.4 wt% to 7.7 wt%, to the PEGDA matrix increases the compliance of the gel. The values of the effective Young moduli calculated for low strains decrease from 0.75 MPa to 0.05 MPa when the PEG-300 000 g.mol$^{-1}$ percentage increases from 0.4 to 7.7 wt%, as represented in Figure 4 b.

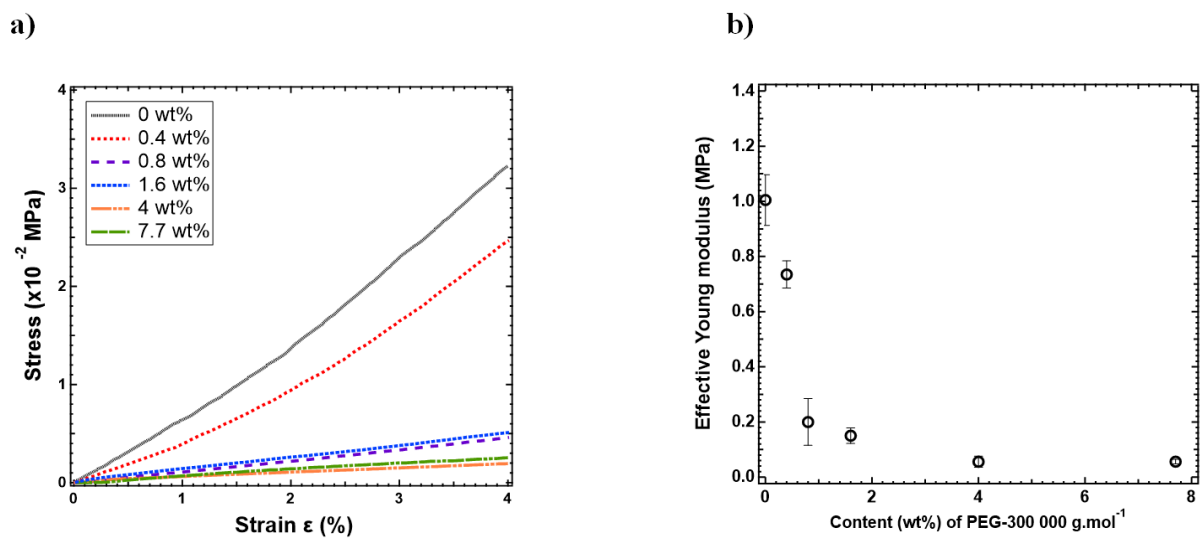

*Figure 4 : a) Stress versus strain for PEGDA hydrogel membranes prepared with 16 wt% of PEGDA and various PEG-300 000 g.mol$^{-1}$ contents in the prepolymerization mixture. b) Variation of the effective Young modulus as a function of PEG-300 000 g.mol$^{-1}$ content.*

During the compression experiments, water is expulsed from the PEGDA/PEG samples (see video in SI 2) while almost no water is expelled in the case of pure PEGDA hydrogels. The fast transport of water expelled from the PEG/PEGDA hydrogels suggests the formation of large and interconnected pores which leads to a larger compressibility and thus lower effective Young moduli in contrast with pure PEGDA samples. The values of the shear modulus G' (presented in SI 3) decrease monotonically from 0.06 MPa to 0.014 MPa when the PEG-300 000 g.mol$^{-1}$ content increases from 0.4 to 4 wt%.

### 3.3. Roles of pressure and PEG concentration on water intrinsic permeability

Filtration experiments were conducted to investigate water intrinsic permeability properties of PEGDA hydrogels membranes prepared with various contents of PEG-300 000 g.mol$^{-1}$. Figure 5 shows the results of the water filtration experiments through a hydrogel membrane composed of 16 wt% of PEGDA which does not contain any free PEG chains. We first notice that the volume of water recovered at constant pressure increases linearly with time, during the filtration process (see Figure 5 a). From the slope of the line, we can deduce the value of the water flow rate which remains constant over time at a given pressure, and increases from $3.26 \times 10^{-12}$ m$^3$.s$^{-1}$ to $1.36 \times 10^{-10}$ m$^3$.s$^{-1}$ when the pressure increases from 2000 Pa to 80000 Pa.

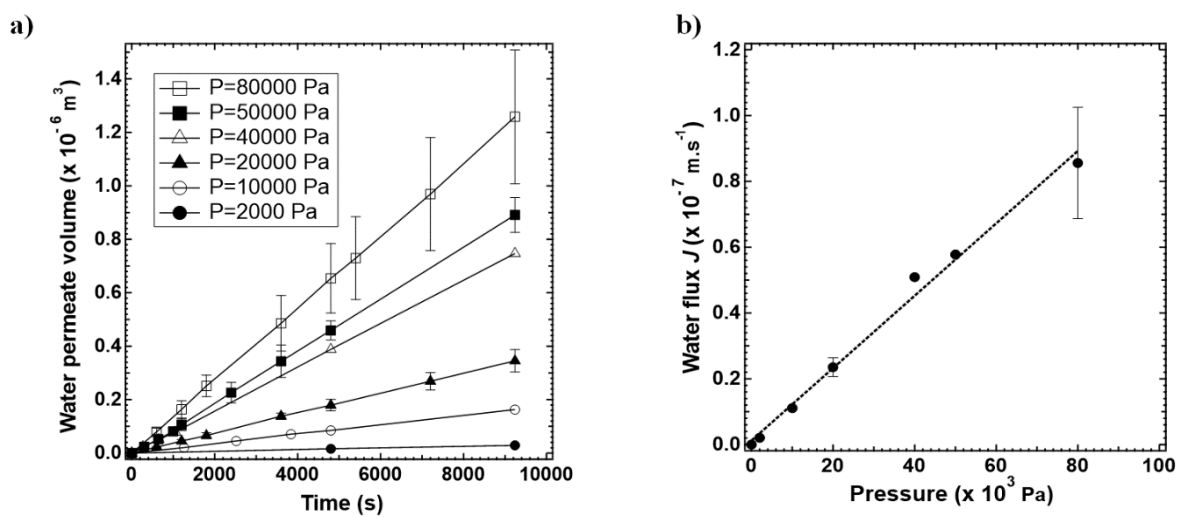

*Figure 5 : a) Water permeate volume variation of PEGDA hydrogel membranes prepared with 16 wt% of PEGDA as a function of time for different pressures. (●) 2000 Pa; (○) 10000 Pa; (▲) 20000 Pa; (△) 40000 Pa ;(■) 50000 Pa and (□) 80000 Pa. b) Water flux variation of a PEGDA hydrogel membrane prepared with 16 wt% of PEGDA as a function of pressure.*

We plot the water flux $J$ as a function of pressure in Figure 5 b. We obtain a linear variation between $J$ and the pressure. The slope of this line enables us to deduce the water intrinsic permeability $K$ using equation (1). We find that $K \sim 10^{-18}$ m$^2$ which is of the same order of magnitude as the values found by Ju and co-workers for a membrane of PEGDA-700 g.mol$^{-1}$ [19].

In the presence of free PEG chains with a molar mass of 300 000 g.mol$^{-1}$ at 1.6 wt%, we first notice that the volume of water recovered at constant pressure increases linearly with time, during the filtration process (see Figure 6 a) as in the previous case of PEGDA hydrogels. Thus, we obtain a constant value of the water flux, $J$, at a constant pressure. As shown in Figure 6 b the water flux for these PEG/PEGDA hydrogels is two orders of magnitude higher than the one obtained for a PEGDA membrane. Interestingly, in contrast to the PEGDA hydrogel, we observe a non-linear variation of the water flux as a function of the applied pressure, as shown in Figure 6 b, and the water flux reaches a plateau at pressures above ~400 mbar. We calculate the intrinsic permeability value for each pressure by using Equation (1). As shown in Figure 6 c, the value of the water intrinsic permeability $K$ decreases by a factor five when increasing the applied pressure by a factor ~40. We will discuss this effect in Section 4.

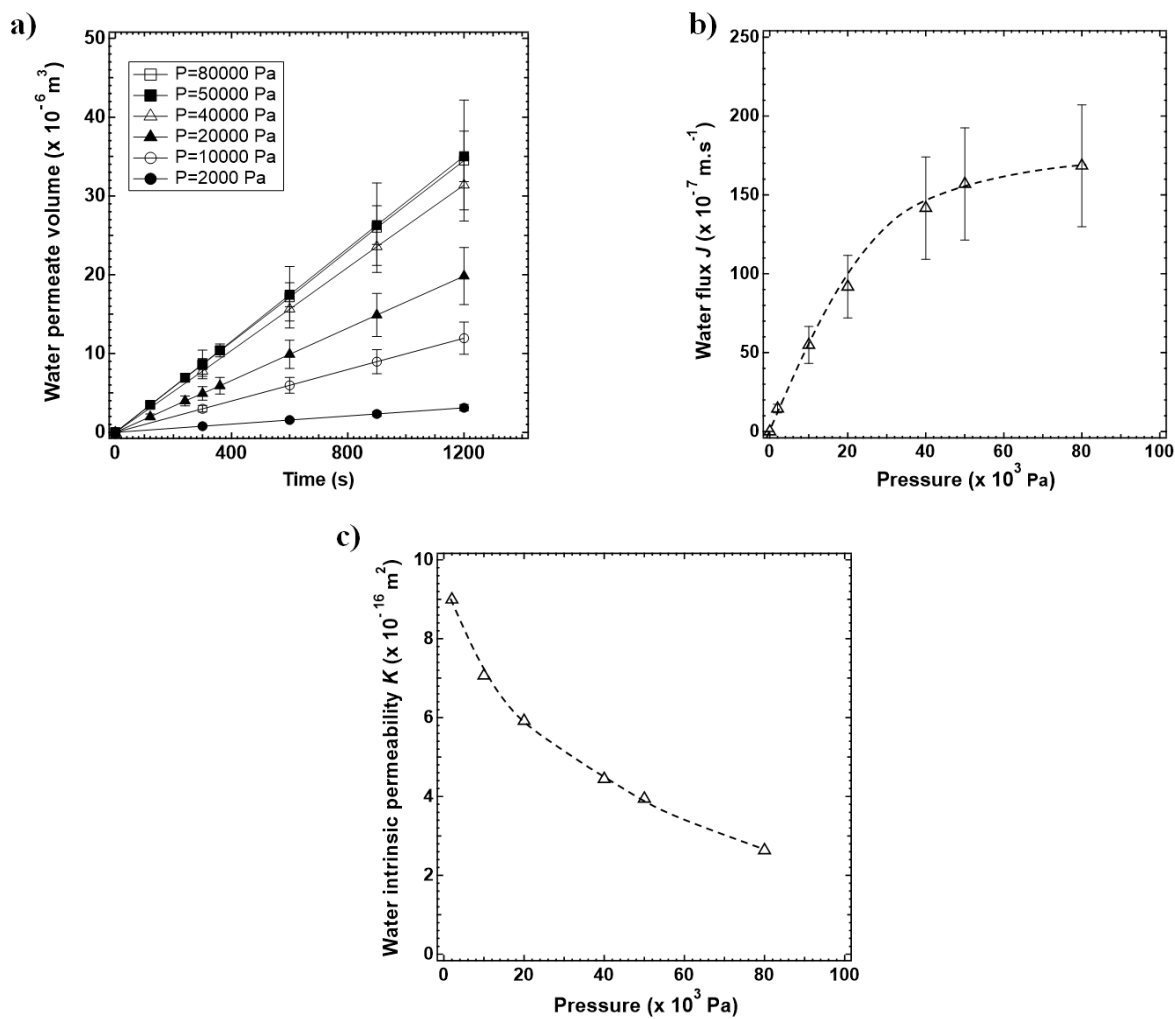

*Figure 6 : a) Water permeate volume variation of PEGDA hydrogel membranes prepared with 16 wt% of PEGDA and 1.6 % of PEG-300 000 g.mol$^{-1}$ as a function of time for different pressures. b) Water flux and c) Water permeability versus pressure for PEGDA hydrogel membranes prepared with 16 wt% of PEGDA and 1.6 wt% of PEG-300 000 g.mol$^{-1}$ contents in the prepolymerization mixture.*
*The dashed lines are represented as guides for the eyes.*

We tested the robustness and reproducibility of the results by performing up to ten filtration cycles between 0 and 1 bar rising and decreasing the pressure. Within each cycle, the water flux-pressure dependence was similar within a 25% deviation, as shown in (Figure 7 a).

By analyzing the PEG content of the permeate after one filtration cycle by total organic carbon (TOC) analysis, we detect a small portion of PEG in the permeate. From the obtained concentration equal to 63 mg.L$^{-1}$ in the permeate, we may estimate the fraction of the PEG that was washed out from the hydrogels, knowing the total quantity of PEG in the hydrogel disks. The fraction therefore reads $f = \dfrac{C_{permeate} * V_{permeate}}{\Phi_{PEG} * \rho_{hydrogel} * V_{hydrogel}}$, with $C_{permeate}$ (mg.L$^{-1}$) the concentration in the permeate measured by TOC, $V_{permeate}$ (L) the volume of permeate, $\Phi_{PEG}$ (wt%) the weight fraction of PEG in the hydrogel, $\rho_{hydrogel}$ the volumic mass of the hydrogel taken equal to $\rho_{water}$ = 10$^6$ mg.L$^{-1}$ and $V_{hydrogel}$ the volume of the hydrogel membrane (L).

The fraction $f$ is about 3% for the first filtration cycle which shows that most of the PEG remains in the hydrogel. For the nine following filtration cycles, the amount of free PEG chains rinsed out of the hydrogel becomes lower than 0.3 % as represented in Figure 7 b.

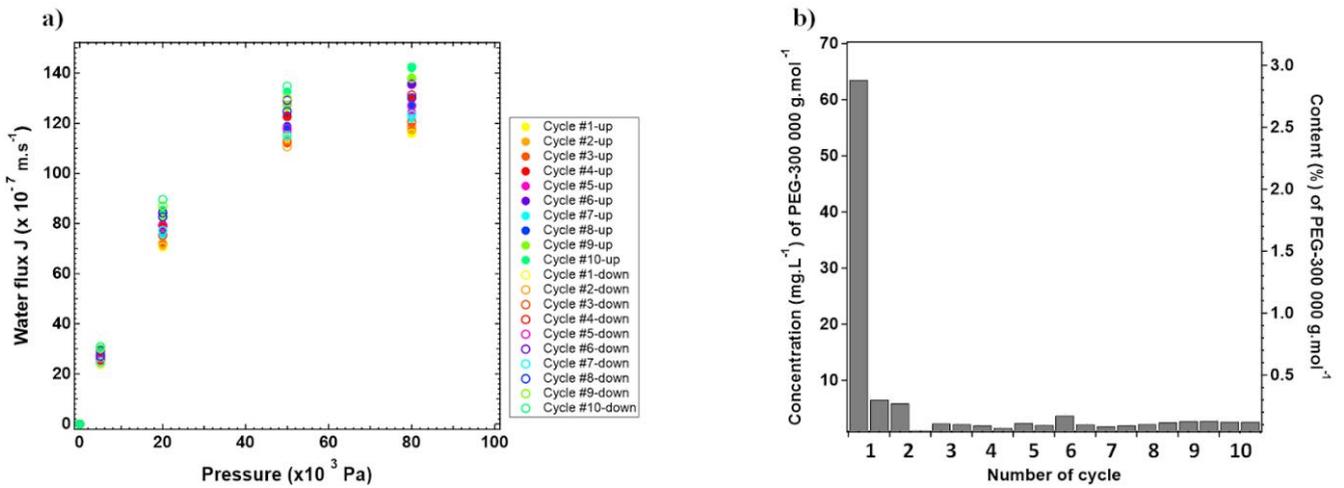

*Figure 7 : a) Variation of the water flux for PEGDA hydrogel membranes prepared with 16 wt% of PEGDA and 1.6 wt% of PEG-300 000 g.mol$^{-1}$ as a function of the applied pressure during ten filtration cycles. b) Effect of the number of filtration cycles on the concentration and the content of PEG-300 000 g.mol$^{-1}$ measured in the permeate volume. The right axis represents the ratio of PEG leaving the hydrogel membrane to the total quantity contained in the membrane.*

These results show that the high molecular weight PEG chains are not rinsed out of the hydrogel during filtration, in contrast to what the data from the literature suggest for short PEG chains that act as templates during the polymerization reaction and get rinsed away afterwards [31-33]. We may attribute this difference to entanglements between the PEG chains and the PEGDA matrix, that resist to the chain withdrawal.

Figure 8 a shows the comparison of the measured water fluxes for PEGDA/PEG hydrogel membranes as a function of pressure at different contents of PEG-300 000 g.mol$^{-1}$. When the PEG percentage increases, the values of *J* exhibit a maximum for a PEG weight fraction of 1.6 wt%. Moreover, for all PEG concentrations, we find a non-linear relation between the water flux and the applied pressure. In Figure 8 b, we plot the water permeability calculated at P=100 mbar according to Equation 1. We observe that *K* varies over two orders of magnitude with the PEG concentration and presents a maximum for a PEG concentration of 1.6 wt%. The possible origins of this behavior will be discussed in the next Section.

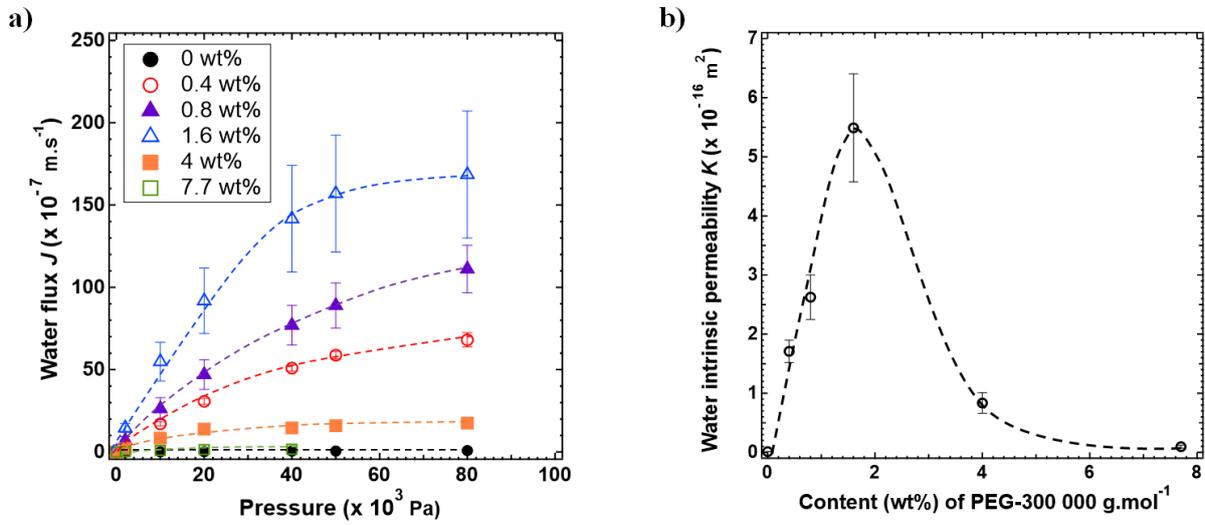

*Figure 8 : Water flux versus pressure for PEGDA hydrogel membranes prepared with 16 wt% of PEGDA and various PEG-300 000 g.mol$^{-1}$ contents in the prepolymerization mixture . (●) 0 wt%; (○) 0.4 wt%; (▲) 0.8 wt%; (△) 1.6 wt% ;(■) 4 wt% and (□) 7.7 wt%. b) Water intrinsic permeability at P=10000 Pa versus content of PEG-300 000 g.mol$^{-1}$ in the prepolymerization mixture of hydrogel membranes prepared with 16 wt% of PEGDA.*
*The dashed lines are guides for the eyes.*

In summary of the permeability measurements, we have shown that addition of free PEG chains to the PEGDA hydrogel network strongly modifies its permeation properties to water. The pressure dependence of the water flux *J* for PEGDA/PEG hydrogels is non-linear, showing a saturation-like behavior at high pressures. The low-pressure permeability strongly varies with PEG concentration and presents a maximum at about 1.6 wt% of free PEG concentration.

## 4. Discussion

### 4.1. Permeability variation with the PEG concentration

As for any porous material, the intrinsic permeability $K$ of a membrane should depend on the pore size $d_p$, the tortuosity $\tau$ and the fraction of open pores [34]. Moreover in the case of a hydrogel, the permeability and kinetics of water transport is known to decrease with the polymer concentration [35]. The permeability of our PEDGA/PEG hydrogels is therefore related to the structure of the hydrogels, which we discuss below.

We observe a maximum in permeability at a PEG concentration of 1.6 wt% which corresponds to the critical overlap concentration $C^*$ of PEG 300 000 g.mol$^{-1}$ chains, as measured by a rheological measurement of the PEG solutions and PEG/PEGDA prepolymerization solutions (see SI 4). Note that the same value of $C^*$ has been obtained by the following relation (3) [36]:

$$(3)\ C^* = \frac{\overline{Mw}}{\frac{4}{3}\pi r_g^3 N_A}$$

where $\overline{Mw}$ is the average molecular weight of the polymer, $r_g$ is the gyration radius (of the order of 20 nm for PEG-300 000 g.mol$^{-1}$) of polymer coils and $N_A$ is Avogadro number: $N_A = 6.023 \times 10^{23}$ mol$^{-1}$.

At concentrations below $C^*$, the AFM measurements performed on the PEGDA/PEG samples showed micron sized zones containing 40 nm large water voids which are smaller than the 200 nm voids observed in PEGDA/water samples. Based on Molina's article which showed that increasing the PEGDA content in PEGDA/water systems leads to a decrease of the void volume fraction, we hypothesize that PEGDA rich micron-sized zones coexist with zones that have a lower PEGDA concentration and may consequently be enriched in PEG. The PEGDA poor phases, which are more permeable than the PEGDA-rich zones, would then lead to an increase of the permeability of the hydrogel membranes.

Above $C^*$, the permeability drops with the PEG concentration. We suggest that the addition of PEG above $C^*$ leads to an increase of the PEG concentration in the PEGDA-poor areas above the entanglement concentration, which may retard the water transportation due to the polymer-solvent friction, similarly to the work reported by Fujiki *et al.* [37] who evidenced that the friction coefficient $f$ -corresponding to $\frac{\mu}{K}$ in our study- is governed by the polymer concentration.

### 4.2. Non-linear variation of the flow rate-pressure curve

To account for the non-linear variation of the flow rate with the applied pressure, we suggest that the membrane compression under the action of a high pressure may lead to the expulsion of water and compression of the pores resulting in a decrease in the gel permeability. The solvent release from hydrogels under compression is a well-known phenomenon and has been studied in detail by Vervoort *et al*. [38]. They demonstrated that under uniaxial compression test, gel deformation, volume loss and solvent expulsion can be observed. This simple experiment shows that the PEGDA/PEG hydrogel structure presents some compressibility.

Assuming that the hydrogel compression is responsible for the permeability loss of the hydrogel with increasing pressure, we expect that the permeability loss should increase with the PEG concentration. In Figure 9, we plot the relative variation of permeability, $[K_{max}-K]/K_{max}$, where $K_{max}$ is the maximum intrinsic permeability obtained at low applied pressure (i.e. P=2000 Pa), as a function of the pressure for several PEG concentrations. The relative variation of the permeability with pressure increases with the PEG concentration, which correlates well with the fact that the hydrogel tends to be more easily deformed when the PEG content increases in the PEGDA hydrogel.

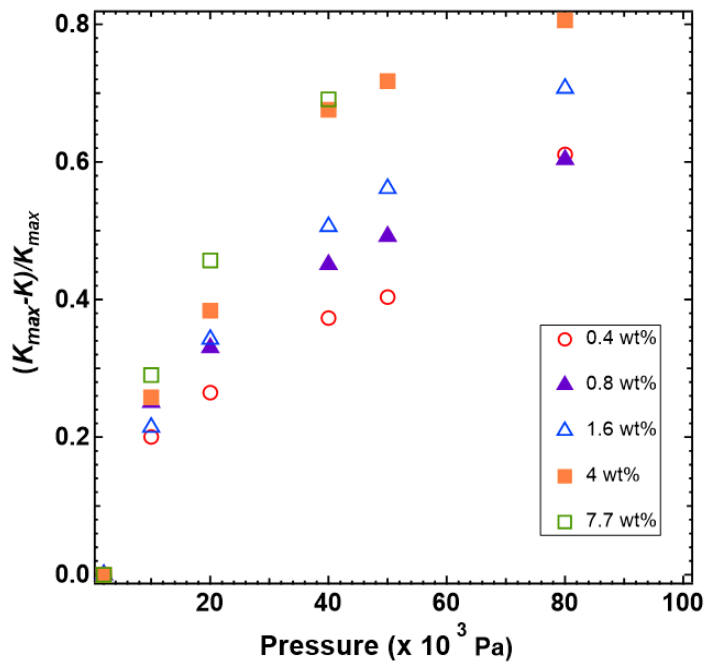

*Figure 9: Water permeability variation versus pressure for PEGDA hydrogel membranes prepared with 16 wt% of PEGDA and various PEG-300 000 g.mol$^{-1}$ contents in the prepolymerization mixture: (○) 0.4 wt%; (▲) 0.8 wt%; (△) 1.6 wt% ;(■) 4 wt% and (□) 7.7 wt%.*

Assuming that the permeability loss is due to the hydrogel compression under the action of the applied pressure difference across the membrane during filtration, we can estimate the variation of the pore size corresponding to the permeability variation. Taking $K \sim d_p^2$ with $d_p$ the pore size of the membrane, according to the Hagen-Poiseuille law, the relative compression of the pores should scale as $\Delta d_p/d_p \sim (1/2)\Delta K/K_{max}$. From the permeability variations reported in Figure 9 we can estimate the range of pore-size variation to be $\Delta d_p/d_p \sim 0.1 - 0.4$ depending on the applied pressure and PEG content in the hydrogel.

The pore compression can also be estimated from the effective Young's modulus E and the pressure as $\Delta d_p/d_p \sim (1 - 2\nu) P/E$, with $\nu$ the effective Poisson's ratio (see details in SI 5). We take $\nu$=0.2 according to Cappello *et al.* who measured the Poisson's ratio of PEGDA/PEG-1000 g.mol$^{-1}$ hydrogels [39].

For a pressure of the order of $10^4$ Pa and taking E ~ 0.1 to 1 MPa, we find a pore variation of $\Delta d_p/d_p \sim$ 0.02 to 0.2 which is, given the rough scaling approach used here, in reasonable agreement with the values of $\Delta d_p/d_p$ estimated above from the permeability variations. This simple argument tells us that the non-linear variation of the permeability with the pressure seems to be mostly due to the compression of the hydrogels and the reduction of the pore size, an effect which increases with the PEG content as these hydrogels have a lower effective modulus than PEGDA hydrogels without PEG.

## 5. Conclusion

We synthesized a PEGDA/PEG composite hydrogel membrane by introducing free PEG-300 000 g.mol$^{-1}$ chains to a PEGDA matrix. FTIR measurement showed that the PEGDA polymerization reaction still occurs even after adding free PEG-300 000 g.mol$^{-1}$ chains of different concentrations. The water flux recovery rate after the cyclic filtration experiments shows that the PEG chains are not flushed out of the gel and thus do not behave as a porogen agent.

Water permeability studies confirmed the validity of Darcy's law for the conventional PEGDA membranes, whereas this law is no longer valid for the PEGDA/PEG composite hydrogels. We assumed in this case that the pressure-induced compression of the hydrogel induces the closure of some pores at high pressure during filtration experiments. Furthermore, increasing the content of PEG chains in the hydrogel system, allows tuning the water permeability over 2 orders of magnitude. The maximum of water permeability is obtained with a hydrogel composite composed of 1.6 wt% of PEG corresponding to the critical overlap concentration $C^*$ of PEG-300 000 g.mol$^{-1}$. Combining these results, as well as turbidity observations and AFM measurements, we hypothesize that the hydrogel structure is controlled by a phase separation between PEGDA rich and PEGDA poor zones, the latter being more permeable to water than the PEGDA rich areas. Below $C^*$, we suggest that the increase of the permeability is obtained because of an increase of the fraction of connected pores in which the water can be transported. Above $C^*$, we suggest that the increase of the PEG concentration in the PEGDA-poor pores may delay the transport of water in the hydrogels leading to a decrease of the permeability.

**Acknowledgements**

We thank A. Marcelan, J. Comtet, G.Ducouret, L. Bocquet and E. Barthel for many fruitful discussions. We gratefully acknowledge Institut Carnot for microfluidics for the financial support during this research project, as well as the Agence Nationale de la Recherche (grants ANR-21-ERCC-0010-01 *EMetBrown,* ANR-21-CE06-0029 *Softer,* ANR-21-CE06-0039 *Fricolas*). We also thank the Soft Matter Collaborative Research Unit, Frontier Research Center for Advanced Material and Life Science, Faculty of Advanced Life Science at Hokkaido University, Sapporo, Japan.

**Supporting Information**

1. **Swelling and water content**

In order to determine the water content in PEGDA and PEGDA/PEG hydrogels, the weight of sample membranes of 1 mm of thickness and 45 mm of diameter is measured right after preparation ($W_0$) and then after placed in an oven at 80°C for several hours ($W_d$). The water percentage of the hydrogels is expressed by the equation (S1):

$$W = \frac{W_0 - W_d}{W_0} \quad (S1)$$

We note that we measure the water content right after polymerization and 24 hours later after immersion in water at room temperature.

Disks were weighed and placed in water at room temperature and then taken out of the solution at time intervals, blotted for removal of the surface water and weighed. The swelling of the network can be expressed by the weight swelling ratio Q in equation (S2):

$$Q = \frac{W_s}{W_0} \quad (S2)$$

where $W_s$ is the weight of the swollen hydrogel.

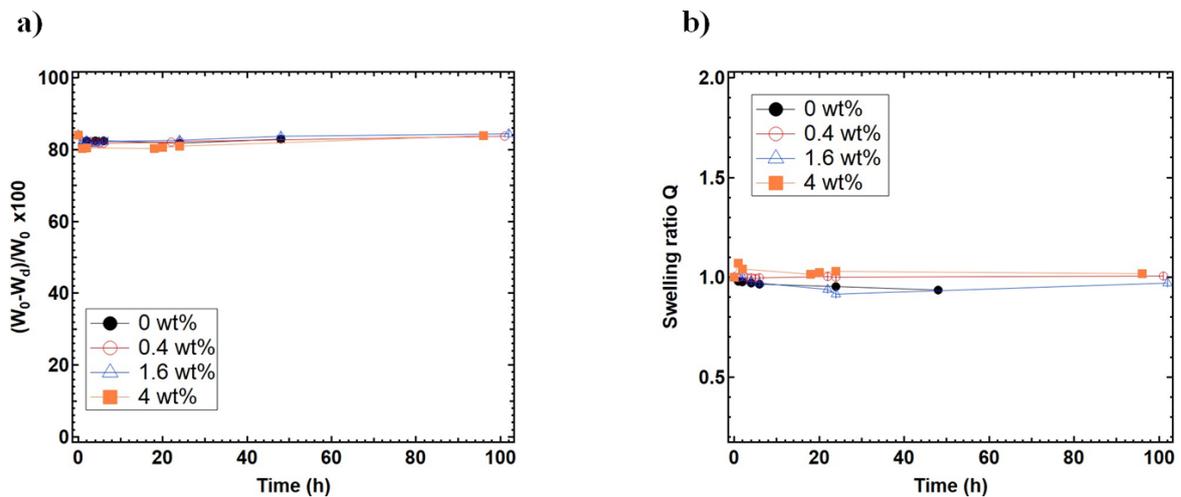

*Figure SI 1: a) Water content and b) Swelling behavior of PEGDA hydrogels membranes prepared with 16 wt% of PEGDA and various PEG-300 000 g.mol$^{-1}$ content in prepolymerization mixture.*

The increasing addition of PEG-300 000 g.mol$^{-1}$ content in prepolymerization mixture does not affect the water content and the swelling behavior of the prepared membrane.

As shown in figure SI 1a, the value of 82 wt% of water content is recovered in the different hydrogels membrane prepared with 0 wt%; 0.4 wt%; 1.6 wt% and 4 wt% of PEG-300 000 g.mol$^{-1}$, after 1 hour of drying at 80°C. This content remains constant even after days of drying.

The swelling behavior of the prepared membrane is represented in figure SI 1b. The results show that conventional and PEG-modified PEGDA membranes hydrogel systems have a swelling ratio close to 1. This means that gels prepared by 16 wt% of PEGDA with or without the addition of PEG-300 000 g.mol$^{-1}$ are in their equilibrium state.

In summary, water content and swelling ratio values of crosslinked membranes are never dependent on the content of free PEG chains in the prepolymerization solution.

### 2. Compression experiments

The high strain imposed on the hydrogel membrane (e.g ~ 30 %), may lead to the expulsion of water from the hydrogel membrane containing a large content of PEG-300 000 g.mol$^{-1}$ chains. For example, the video SI 2 a) shows that as soon as the stress is increased (loading phase), the PEGDA/PEG-1.6 wt% gel diameter starts to increase and water is expelled. During the unloading phase, we observe that the hydrogel recovers its initial shape and reabsorbs again the water that was expelled. Whereas, for PEGDA/PEG-0 wt% hydrogel, the second video SI 2 b) shows an increasing of the gel diameter but without the expulsion of water during the compression test. This behavior can be explained by the formation of large pores in the PEGDA/PEG hydrogel membrane.

### 3. Shear rheology measurements

The viscoelastic properties of a series of PEGDA/PEG hydrogel membranes were investigated by oscillatory measurements on a TA-DH3 rheometer instrument equipped with a plate-plate geometry (diameter: 20 mm) and solvent trap. A strain sweep test was conducted under an angular frequency of 6.27 rad. s$^{-1}$. The initial axial force $F_0$ applied is 0.4 N. The results represented in Figure SI 3 a show the variation of the shear modulus as a function of the oscillation strain for the hydrogels membranes prepared with various PEG-300 000 g.mol$^{-1}$ content.

In the linear viscoelastic regime, the shear modulus G' is independent of the oscillation deformation, so that a plateau is obtained. On the other hand, the decrease in the value of the shear modulus at high strain signifies the transition to the non-linear regime. As soon as we increase the PEG content in the hydrogel membrane, the linear regime is more extended.

The value of shear modulus G' obtained at the plateau as a function of the PEG-300 000 g.mol$^{-1}$ content is represented in Figure SI 3 b. The addition of free PEG chains, from 0.4 wt% to 4 wt% decreases the shear modulus values from 0.06 MPa to 0.014 MPa, respectively.

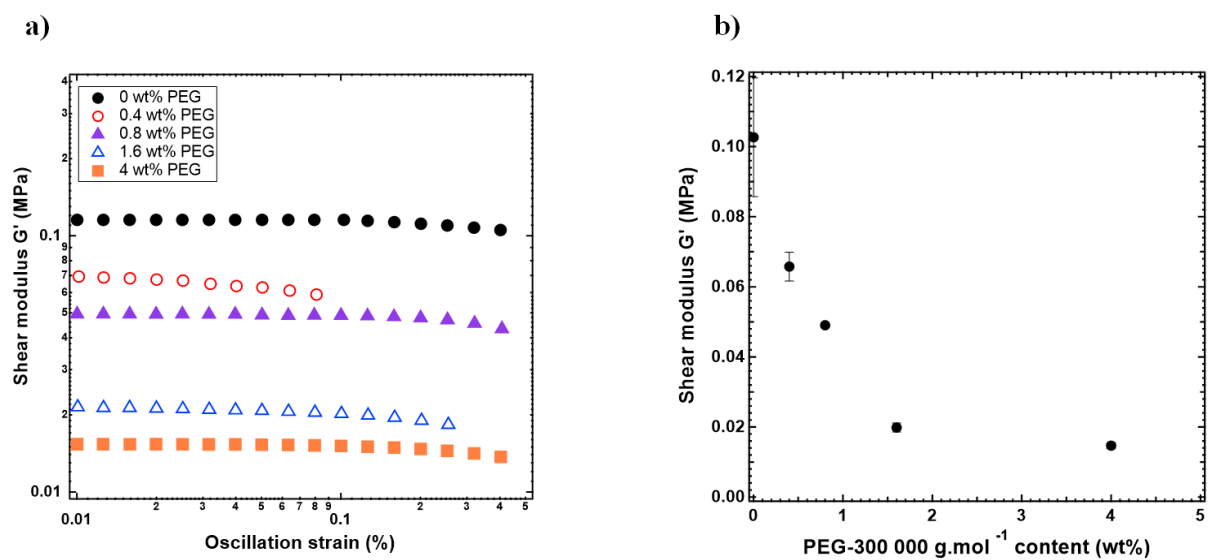

*Figure SI 3: a) Variation of shear modulus as a function of oscillation strain at $F_0$=0.4 N for the various PEG-300 000 g.mol$^{-1}$ content. b) Variation of shear modulus as a function of the various PEG-300 000 g.mol$^{-1}$ content.*

### 4. Rheological experiments

Rheological experiments are done using an ARG-2 rheometer TA, in order to obtain the critical overlap concentration $C^*$ of PEG $\overline{Mw}$=300 000 g.mol$^{-1}$ chains.

As shown if Figure SI 4, as expected the viscosity increases when the PEG concentration increases. At some point, there is a transition between the diluted and the semi-diluted regime, represented by the variation of the slope of the line slope from 0.96 to 3.2, respectively. At this stage, there is a contact between the PEG coils due to the decrease of the distance between them. The associated concentration is called critical overlap concentration $C^*$ being determined at 1.6 wt% of PEG.

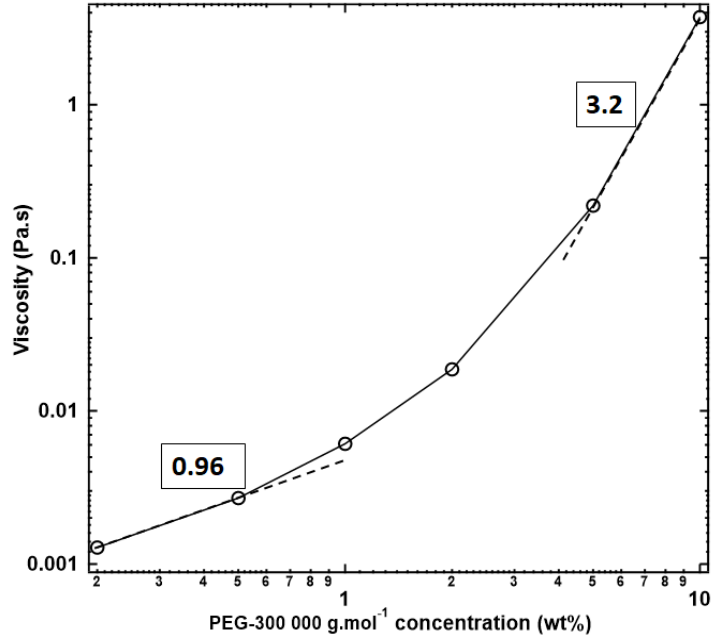

*Figure SI 4: The variation of the viscosity as a function of various PEG-300 000 g.mol$^{-1}$ concentration.*

## 5. Model of pore compression

Let us take a composite material made of an incompressible elastic solid matrix (PEGDA) of volume $V_0$ and Young's modulus $E_0$, and a set of $N_p$ identical liquid pores (or inclusions) of size $d_p$, or volume $Vp \sim d_p^3$. The ensemble has an effective Young's modulus $E(\varphi)$, where $\varphi = Np.Vp/V$ is the volume fraction of liquid (or of pores), and where $V = Vo + Np.Vp$ is the total volume. We also introduce an effective Poisson's ratio $\nu < 0.5$. When this matrix full of liquid pores is compressed isotropically with a pressure $P$, it is the effective compression/bulk modulus B, and not the effective Young's modulus E which it is necessary to take into account. The elastic relation in this case is $P = B.dV/V$, where $dV$ is the variation of the total volume. The two modules are linked by the law $B = \dfrac{E}{3(1-2\nu)}$. However, one has $dV = Np.dVp$ since $V_o$ is constant due to the presumed incompressibility of the matrix. Thus, $P = B.N_p.\dfrac{dV_p}{V} = B.\varphi.\dfrac{dV_p}{V_p} = 3.E.\varphi.\dfrac{dd_p}{d_p} = \dfrac{\varphi.E}{(1-2\nu)}\dfrac{dd_p}{d_p}$. Finally, one needs to have an expression for $E(\varphi)$. In a work by Style *et al.* [40] for example, a model for identical spherical inclusions is provided. In their work, we see that for fractions that are not too low, the product $\varphi.E(\varphi)$ is of the order of $E_0$, whatever the elastocapillary ratio $\gamma/(E_0.d_p)$, with $\gamma$ the solid-liquid surface tension. Hence, modeling liquid pores as these inclusions, one gets in fine: $P \sim \dfrac{E_0}{(1-2\nu)}\dfrac{dd_p}{d_p}$ as a reasonable approximation.